\documentclass[aps,prd,twocolumn,superscriptaddress,amsmath]{revtex4-1}

\usepackage{epsf}
\usepackage{graphicx}
\usepackage{latexsym}

\usepackage[normalem]{ulem}

\usepackage{color}

\newcommand{\be}{\begin{equation}}
\newcommand{\ee}{\end{equation}}
\newcommand{\bea}{\begin{eqnarray}}
\newcommand{\eea}{\end{eqnarray}}

\newcommand{\sumint}{\int\!\!\!\!\!\!\!\!\!\sum}
\DeclareMathOperator{\Tr}{Tr}
\DeclareMathOperator{\Ei}{Ei}
\DeclareMathOperator{\realpart}{\Re e}
\DeclareMathOperator{\imagpart}{\Im m}

\begin{document}


\preprint{APS/123-QED}

\title{Casimir effect in free-fall towards a Schwarzschild black hole}

\author{Francesco Sorge}
 \email{sorge@na.infn.it}
\affiliation{I.N.F.N. - Complesso Universitario Monte S. Angelo, via Cintia, Ed. 6,
80126 Napoli, Italy
}%


\author{Justin H. Wilson}
 \email{justin@jhwilson.com}
\affiliation{%
Institute of Quantum Information and Matter and Department of Physics,
Caltech, CA
}%
\affiliation{
Department of Physics and Astronomy, Center for Materials Theory, Rutgers University, Piscataway, NJ 08854 USA
}



\date{\today}


\begin{abstract}
In this paper we discuss the Casimir effect in a small cavity, freely falling from spatial infinity in spacetime geometry outside of a Schwarzschild black hole.
Our main goal is to search for possible changes in the vacuum energy, as well as particle creation inside the falling cavity, with respect to a comoving observer.
Working in the Lema\^itre chart and assuming a cavity size $L$ much smaller than the Schwarzschild radius ($L/r_g\ll 1$), we solve the Klein-Gordon equation for a massless scalar field confined within the cavity in the reference frame of the comoving observer.
We follow Schwinger's proper time approach, evaluating the one-loop effective action for the field in the falling cavity hence evaluating the corrections to the vacuum energy.
We find a small reduction in the absolute value of Casimir energy as the cavity approaches the black hole horizon due to the changing spacetime geometry.
Since the spacetime geometry for the cavity changes dynamically, we further find the energy density of the created particles due to the dynamical Casimir effect.
These dynamical contributions exactly match the deficit to the static Casimir energy.
Combined, the obersever measures a net increase in energy within the cavity as she falls.
\end{abstract}



\maketitle


\section{Introduction}


The Casimir effect \cite{casimir1,casimir2} is one of the most intriguing aspects of quantum field theory (QFT) where the energy of the vacuum gives rise to a measurable \cite{experiment} force between macroscopic objects.
Roughly speaking, it originates from a distortion in the modes of a quantum field constrained in a finite region of space by some boundaries.
This distortion can arise from material properties \cite{lifshitz} as well as from the background spacetime's geometry \cite{casimir3}.
In the latter case, the Casimir effect becomes an exciting arena in which general relativity (GR) and quantum field theory (QFT) face each other.
Indeed, the Casimir effect in presence of gravito-inertial fieds has been considered in detail by many authors through the years, giving rise to a rich literature concerning the issue \cite{calloni1,calloni2,marquez1,marquez2,marquez3,marquez4,sorge1,milton,fulling1,sorge2,sorge3}.

When the background spacetime geometry is time-varying, we are faced with further dynamical effects, typically related to particle creation out of the quantum field vacuum \cite{Moore,parker1,parker2,fulling2,davies,unruh,birrell}.
So a Casimir cavity becomes an interesting {\em laboratory}, where both vacuum polarization and vacuum persistence can be explored in detail.
QFT generally relies upon a partitioning of space-time in time-like surface (Cauchy surfaces) upon which to build a Hamiltonian.
However, general relativity has a frame independence built into it, and the tension between these two theories leads to many interesting effects including Hawking radiation \cite{hawking}.
This tension becomes most relevant when strong gravito-inertial regimes are met.
Because of the intrinsic weakness of gravity, the most favourable conditions are those involving highly collapsed massive bodies, as black holes.

The influence of a gravito-inertial environment on a Casimir cavity can give rise to several changes in the vacuum energy. According to their origin, we may consider
\begin{itemize}
\item{{\em tidal} effects: due to the spatial extension of the Casimir apparatus, these are expected to cause anisotropies in the distribution of the vacuum energy density inside the cavity. Such effects have been discussed in \cite{justin}, where a detailed analysis of 1+1D model of Casimir cavity falling into a Schwarzschild black hole has been extensively performed.}
\item{{\em pure geometric} effects: also when tidal effects are neglected, we can still face possible corrections to Casimir energy due to the change in spacetime geometry.
In particular, the quantum fields probe a finite extent of spacetime and can therefore be sensitive to the geometry's variation in time [as captured by the modified Klein-Gordon equation we derive later in Eq.~\eqref{KG}].
The stress-tensor is sensitive to this as well, and therefore local measurements performed by an observer could witness such changes.
}
\end{itemize}

In this paper we will focus on the \emph{pure geometric} effects.
Both the above cited effects are part of the same overall effect and could even apply at the same order in perturbation theory, but we suspect they contribute additively and can thus be separated \footnote{Since everything in this work is reduced to a local metric that varies with the observer's proper time $\tau$, one could call the ``pure geometric'' effect a ``time-tidal'' effect since it probes variation in the gravitational field in time instead of across space like the usual tidal effect.}.
Indeed, even the latter effect appears similar to a tidal effect insofar as a local observer can use it as a probe for if she is in a gravitational field---much akin to the use of the classical tidal effect to determine such.
We leave the analysis of the 3+1D tidal effects to future work.

Our starting point is a small Casimir cavity freely falling (from spatial infinity) in the gravitational field of a Schwarzschild black hole.
We assume that the typical cavity size is much smaller than the black hole gravitational radius, $r_g =2M$. In particular, this means that $L\ll r_g$, where $L$ is the proper plate separation.

We are interested in possible changes in the vacuum energy density detected by an observer comoving with the cavity.
We could anticipate that such changes - if any - will be likely to appear near the black hole horizon where the Schwarzschild metric has a coordinate singularity.
To avoid such an obstacle, we will employ the Lema\^itre chart \cite{lemaitre,kramer} which has the advantage of being regular at the horizon; further, it will be especially  useful in describing freely falling observers, as we will see below.

Here are the basic steps.
First, we solve (in the observer's local frame) the Klein-Gordon equation for a massless, minimal coupled, scalar field inside the cavity.
Subsequently we use Schwinger's proper time approach in deriving the one-loop effective action for the quantum field.
A pleasant feature of the chosen approach is that it could allow, in priniciple, also for a non-perturbative analysis.
From the effective action we finally deduce the Casimir effect as well as the small (static as well dynamical) corrections to the energy density due to the cavity fall.

The results, although very small as expected, show a tiny change in the Casimir energy.
Namely, we find a quite small reduction (in absolute value) of its flat spacetime static value, $\langle\epsilon_\mathrm{Cas}\rangle_0=-\frac{\pi^2}{1440L^4}$.
We also obtain a small contribution due to particle creation inside the Casimir cavity that happens to match the contribution to the static value.

The paper is organized as follows.
In section~\ref{sec:LemaitreCoordinates}, we review the coordinate transformation yielding the generalized Lema\^itre form of the Schwarzschild spacetime.
We then specialize to the case of a test body freely falling from the spatial infinity with zero initial velocity.
In section~\ref{sec:CasimirComoving}, we introduce the Casimir cavity while also stating the basic assumptions of the model.
Subsequently, we derive the tetrad frame adapted to a physical observer comoving with the cavity.
In section~\ref{sec:ScalarField}, we solve the Klein-Gordon equation for a massless scalar field inside the falling cavity (assuming minimal coupling).
In section~\ref{sec:SchwingerApproach}, we  follow Schwinger's proper-time method \cite{schwinger1,schwinger2,cougo} in order to deduce the one-loop effective action $W$ for the quantum field.
We discuss the real and the imaginary part of $W$, related to the vacuum polarization and vacuum persistence, respectively.
In section~\ref{sec:StaticCasimir} we consider the vacuum polarization, from which we deduce the static Casimir effect as well as the corrections due to the cavity fall.
In section~\ref{sec:DynamicalEffects} we discuss the dynamical aspects, namely particle creation inside the cavity, analyzing the vacuum persistence contribution.
By means of the Bogolubov approach, we evaluate the energy density in terms of created field quanta inside the falling cavity. We discuss the results in section~\ref{sec:Discussion} while section~\ref{sec:DynamicalEffects} is devoted to some final remarks.

Throughout the paper, unless otherwise specified, use has been made of natural geometrized units.
Greek indices take values from 0 to 3; latin ones take values from 1 to 3.
The metric signature is $(+,-,-,-)$, with determinant $g$.

\section{Lema\^itre coordinates: an overview}\label{sec:LemaitreCoordinates}
The Schwarzschild metric for a black hole of mass $M$ in the standard Schwarzschild coordinates $\{t,r,\theta,\phi\}$ reads
\begin{equation}\label{schw}
ds^2= \bigg(1-\frac{r_g}{r}\bigg)dt^2-\bigg(1-\frac{r_g}{r}\bigg)^{-1}dr^2-r^2 d\Omega^2,
\end{equation}
where $r_g=2M$ is the gravitational (Schwarzschild) radius of the black hole and $d\Omega^2=d\theta^2+\sin^2\theta d\phi^2$.
In such coordinates, there is a {\em coordinate} singularity at the horizon.
Being interested in the behavior of a Casimir cavity falling into a black hole, we need a chart which is regular at the horizon, so the form \eqref{schw} of the metric is not suitable.
Among the various coordinate systems that are well-behaved at the horizon, we will adopt the Lema\^itre chart, which will prove useful when describing the free-fall of the Casimir cavity near the horizon.

Curiously, only little work can be found in the literature about Lema\^itre coordinates \cite{kramer}, concerning both their deduction and their practical applications. Therefore, let us briefly recall how the Lema\^itre chart can be obtained from the Schwarzschild coordinates.  Consider a massive test body, radially falling with four-velocity $\bf u$ in the gravitational field of the black hole.

Since Eq.~\eqref{schw} admits a time-like Killing vector, $\vec X=\partial_t$, we have a conserved quantity along a time-like geodesic, namely $\vec X\cdot \bf u=\gamma=$const, with $\gamma={\cal E}/m$ being the total specific energy of the test body (if the test body starts falling from rest at the spatial infinity then $\gamma=1$).
Since for a radial infall motion $d\theta =d\phi=0$, we get from the constraint $g_{\mu\nu}u^\mu u^\nu=1$
\begin{equation}\label{ur}
\frac{dr}{d\tau}=-\sqrt{\gamma^2-1+\frac{r_g}{r}},
\end{equation}
where the sign refers to the radial {\em in fall} and $\tau$ is the proper time of the falling test body.
Notice that Eq.~\eqref{ur} implies
\begin{equation}\label{vincle}
\gamma^2-1+\frac{r_g}{r}>0,
\end{equation}
otherwise we have no radial motion.
Such a constraint defines the allowed radial region as a function of the Schwarzschild radius as well as the total specific energy of the falling body.

From Eq.~\eqref{ur} we formally have
\begin{equation}\label{rint}
-\tau+c=\int\,\frac{dr}{\sqrt{\gamma^2-1+\frac{r_g}{r}}}=F(r;r_g,\gamma),
\end{equation}
where $F(r;r_g,\gamma)$ is a rather cumbersome function, defined as
\begin{widetext}
\begin{equation}
F(r;r_g,\gamma)=\left\{
	\begin{array}{ll}
	|\gamma^2-1|^{-1/2}\bigg[\sqrt{r\bigg(\frac{r_g}{\gamma^2-1}+r\bigg)}-\frac{r_g}{\gamma^2-1}\ln\bigg(\sqrt{r}+\sqrt{\frac{r_g}{\gamma^2-1}+r}\bigg)\bigg], &\gamma\neq 1 \\
	\frac{2}{3}\frac{r^{3/2}}{\sqrt{r_g}}, & \gamma = 1.
	\end{array}
\right.\label{rfunction}
\end{equation}
\end{widetext}
In Eq.~\eqref{rint} $c$ is an arbitrary integration {\em constant}.
Notice that, for any value of $c$, Eq.~\eqref{rint}  describes (although implicitly) a physically admissible time-like geodesic for an infalling test body (recall that $\gamma=1$ means free fall from spatial infinity with zero initial velocity).
This suggests  {\em defining} a new radial coordinate $\rho$ by letting just $\rho=c$.
In so doing, a freely falling body is defined by a constant value of the coordinate $\rho$, hence we write
\begin{equation}\label{rint2}
-\tau+\rho=F(r;r_g,\gamma),
\end{equation}
In other words, we are defining a {\em comoving coordinate, adapted to time-like geodesics}: a body moving along such a geodesic has a proper time $\tau$ and a constant value of the coordinate $\rho$.

From Eq.~\eqref{rint2}, we also get the following relationship between the coordinate differentials
\begin{equation}\label{dr}
dr= \sqrt{\gamma^2-1+\frac{r_g}{r}}(d\rho-d\tau).
\end{equation}
We now search for a similar relation involving the Schwarzschild time $t$. We guess
\begin{equation}\label{dt}
dt=Ad\rho + B d\tau,
\end{equation}
with $A$ and $B$ unknowns to be determined requiring that the Schwarzschild metric in the new coordinates $\{\tau,\rho,\theta,\phi\}$ is adapted to the falling body, namely $g_{\tau\tau}=1$ (syncronous coordinate system) and $g_{\tau\rho}=0$ (diagonal metric).
Substituting  (\ref{dr}) and (\ref{dt}) in (\ref{schw}) we have
\begin{equation} \label{eq:coord-transformation}
\begin{split}
g_{\tau\tau}& =\bigg(1-\frac{r_g}{r}\bigg)B^2- \frac{\gamma^2 - 1 + \frac{r_g}r}{1 - \frac{r_g}r} = 1
\\
g_{\tau\rho}& =\bigg(1-\frac{r_g}{r}\bigg)AB+\frac{\gamma^2 - 1 + \frac{r_g}r}{1 - \frac{r_g}r}=0,
\end{split}
\end{equation}
from which we obtain $A=\frac{1}{\gamma}-\frac{\gamma r}{r-r_g}$ and $B=\frac{\gamma r}{r-r_g}$. Thus, the full required coordinate transformation reads
\begin{equation} \label{transf}
\left\{
	\begin{array}{ll}
dt= \frac{\gamma r}{r-r_g}d\tau+\bigg(\frac{1}{\gamma}-\frac{\gamma r}{r-r_g}\bigg)d\rho\\
dr= \sqrt{\gamma^2-1+\frac{r_g}{r}}(-d\tau+d\rho),
\end{array}
\right.
\end{equation}
or, in matrix form
\begin{equation}
d\vec x_S=Q(\gamma)d\vec x_L,
\end{equation}
where $Q(\gamma)$ is the matrix defined from Eq.~\eqref{transf} and $d\vec x_S=(dt,dr)^T$, $d\vec x_L=(d\tau,d\rho)^T$ are the coordinate 1-forms in the Schwarzschild and Lema\^itre coordinates respectively.
Inverting $Q(\gamma)$, we obtain
\begin{equation}
\left\{
	\begin{array}{ll}
d\tau=\gamma dt+\gamma^2\bigg(1-\frac{r_g}{r}\bigg)^{-1}\bigg(\gamma^2-1+\frac{r_g}{r}\bigg)^{-1/2}\frac{r_g}{r}dr,\\
d\rho=\gamma dt+\gamma^2\bigg(1-\frac{r_g}{r}\bigg)^{-1}\bigg(\gamma^2-1+\frac{r_g}{r}\bigg)^{-1/2}dr.
\end{array}
\right.
\end{equation}
Using Eq.~\eqref{transf} in Eq.~\eqref{schw} yields the Schwarzschild metric in the so-called {\em generalized} Lema\^itre coordinates $\{\tau,\rho,\theta,\phi\}$ (see, e.g., \cite{kramer})
\begin{equation}\label{glm}
ds^2=d\tau^2-\frac{1}{\gamma^2}\bigg(\gamma^2-1+\frac{r_g}{r(\tau,\rho)}\bigg)d\rho^2-r^2(\tau,\rho)d\Omega^2,
\end{equation}
where $r(\tau,\rho)$ is implicitly given by Eq.~\eqref{rint2}. The existence of the inverse function $r(\tau,\rho)$ is assured, since the Jacobian $J$ of the transformation Eq.~\eqref{transf} is $J=\det\,Q(\gamma)=\frac{1}{\gamma}\sqrt{\gamma^2-1+\frac{r_g}{r}}>0$ [recall the constraint Eq.~\eqref{vincle}].

Let us briefly comment about the spacetime symmetries.
Inspection of Eq.~\eqref{schw} immediately tell us that $\vec X=\partial_t$ is a Killing vector for the Schwarzschild spacetime (in the Schwarzschild coordinates) as the metric is independent of $t$.
This time-like Killing vector field implies an energy conservation in the Schwarzschild spacetime.
Although not explicitly visible, such symmetry exists in the Lema\^itre coordinates as well.
The corresponding form of the Killing vector can be obtained from the transformation Eq.~\eqref{transf}, by means of the relationship  between the canonical basis vectors $\partial_S=(\partial_t,\partial_r)^T$ and $\partial_L=(\partial_\tau,\partial_\rho)^T$
\begin{equation}\label{basistransf}
\partial_S=[Q(\gamma)^{-1}]^T\partial_L.
\end{equation}
From Eq.~\eqref{basistransf} we immediately get $\partial_t=\partial_\tau+\partial_\rho$. So, in the Lema\^itre coordinates energy conservation is related to the Killing vector $\vec X=\partial_\tau+\partial_\rho$.

Consider now a freely falling test body, with total specific energy $\gamma$.
Adjust the test body's clock so that the proper time $\tau=0$ occurs when it is at a given radial Schwarzschild coordinate $r_0$.
Putting $r=r_0$ and $\tau=0$ into Eq.~\eqref{rint2} we get the corresponding value of the comoving radial Lema\^itre  coordinate $\rho_0=\rho(0,r_0)$ at the initial proper time $\tau=0$.
Replacing again the {\em constant} value  $\rho=\rho_0=\rho(0,r_0)$ in Eq.~\eqref{rint2} we implicitly get the radial coordinate $r$ as a function of the proper time $\tau$, namely $r=r(\tau;\rho_0)$.

The above procedure is easy to carry out when the test body has $\gamma=1$.
In this case, the Schwarzschild metric in the Lema\^itre coordinates reduces to
\begin{equation}\label{lm}
ds^2=d\tau^2-\frac{r_g}{r(\tau,\rho)}d\rho^2-r^2(\tau,\rho) d\Omega^2.
\end{equation}
From Eqs.~\eqref{rfunction}) and \eqref{rint2}, we immediately get
\begin{equation}\label{r}
r(\tau,\rho)=r_g^{1/3}\bigg[\frac{3}{2}(\rho-\tau)\bigg]^{2/3}.
\end{equation}
As discussed above, for any admissible {\em fixed} value of the radial coordinate $\rho$, Eq.~\eqref{r} describes the radial motion of a test body freely falling from spatial infinity with zero initial velocity ($\gamma=1$).
Being interested in the behavior near the black hole horizon (where the Lema\^itre coordinates are regular), we fix the proper time origin $\tau=0$ just at the horizon crossing $r=r_g$.
From Eq.~\eqref{r} we get the {\em constant} value of the radial coordinate $\rho$ along the corresponding geodesic
\begin{equation}
\rho_0=\frac{2}{3}r_g\quad\quad\quad (\tau=0).
\end{equation}
From Eq.~\eqref{r} we obtain
\begin{equation}\label{r1}
r(\tau;\rho_0)=r_g\bigg(1-\frac{3\tau}{2r_g}\bigg)^{2/3},
\end{equation}
representing a freely falling particle (in our case the Casimir cavity) whose trajectory intersects the horizon at $\tau=0$.
Notice, in passing, that the travel from the infinity to the horizon is described by {\em negative} values of the proper time: $-\infty<\tau\leq 0$.
Also, reaching the singularity from the horizon takes a finite proper time $\tau_s=\frac{2}{3}r_g$.

\begin{figure}
\begin{center}
\includegraphics[angle=0,width=.4\textwidth]{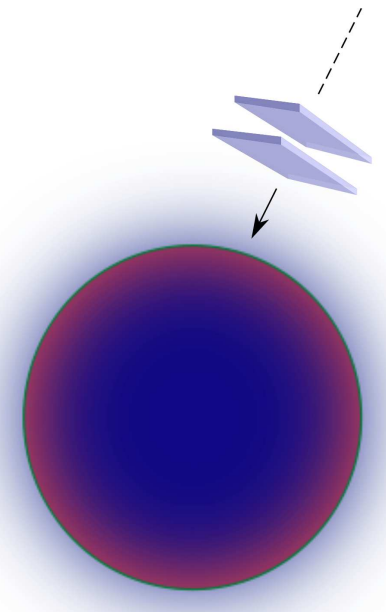}%
\caption{Schematic picture  of  a Casimir cavity falling onto a Schwarzschild black hole.
We assume the cavity is small with respect to the black hole gravitational radius, ($L/r_g\ll1$), falls from spatial infinity with zero initial velocity ($\gamma=1$) and zero angular momentum, and is {\em rigid}, namely the plate separation $L$ is {\em constant} according to a comoving observer.
Lastly, we neglect variations in the gravitational field across the apparatus (see text for details). \label{fig1}}
\end{center}
\end{figure}

\section{The Casimir cavity and the comoving frame} \label{sec:CasimirComoving}
The measurement of Casimir energy inside the falling cavity is performed by a \emph{comoving} observer.
Before proceeding we need some assumptions about the cavity and the reference frame with respect to which the observer makes her measurements.
Concerning the cavity, we take its geometry so that the plates (of area $A$ and separated by a distance $L$, such that $L\ll\sqrt{A}$) are orthogonal to the radial falling direction \footnote{Such a choice has been made only for the sake of definiteness.}. We further require that:
\begin{itemize}
\item{the cavity is taken to fall from spatial infinity with zero initial velocity ($\gamma=1$) and zero angular momentum;}
\item{the typical cavity size is much smaller than the gravitational radius of the black hole, so that, in particular, $L\ll r_g$, with $L$ being the  plate separation;}
\item{the cavity is {\em rigid}; its dimensions and shape do not suffer any distortion, in spite of external tidal forces (imagine a rigid rod, invisible to the scalar field, that holds the plates about its center of mass);}
\item{the center of mass of the cavity follows a true geodesic motion; hence we neglect other non-gravitational external effects, including those possibly related to backreaction;}
\item{the change in the gravitational field across the apparatus at a given proper time is negligible.}
\end{itemize}
We stress that the last assumption is rather subtle and is controlled but at the same level of perturbation theory as the following analysis.
However, corrections due to the last assumption we expect to enter in linearly at the same order in perturbation theory and could thus be isolated, and furthermore, such an effect comes from a qualitatively different source than what we are considering herein (the proper time variation of the spacetime geometry \emph{across} the entire apparatus, not its spatial variations).
A deeper analysis of tidal effects on Casimir energy in a 1+1D falling cavity has been extensively performed in \cite{justin}, and we leave the analysis of the 3+1D case for future work.

As a next step we choose a reference frame, defining a tetrad adapted to the comoving observer.
We will work in the Lema\^itre coordinates.
The metric in Eq.~\eqref{lm} is diagonal and thus, the required tetrad $\{e_{\hat a}^\mu\}$ can be readily obtained.
From Eq.~\eqref{lm} we have (using $a, b, c,...$ to label tetrad indices)
\begin{equation}\label{tetrad}
\begin{split}
e_{\tau}&=\partial_\tau\\
e_{x}&=\sqrt{\frac{r}{r_g}}\partial_\rho\\
e_{y}&=\frac{1}{r}\partial_\theta\\
e_{z}&=\frac{1}{r\sin\theta}\partial_\phi.
\end{split}
\end{equation}
So, the observer performs her measurements in the (locally minkowskian) rectangular coordinates $\{\tau,x,y,z\}$.
In the following, exploiting spherical symmetry, we will put $\theta=\pi/2$.
We also have $e=\sqrt{-g}=r^2\sqrt{\frac{r_g}{r}}$.
It is understood that in Eq.~\eqref{tetrad} $r=r(\tau;\rho_0)$ is given by Eq.~\eqref{r1}.
This is precisely our last assumption in the above list: that the geometry does not change across the apparatus.
Technically, we ought to have $r = r(\tau; \rho)$ where $\rho$ varies between the two curves the plates follow $\rho = \rho_{1,2}(\tau)$ which are technically not geodesics.
With our assumption that $r(\tau;\rho) \approx r(\tau; \rho_0)$, the small variation is neglected and the corrections of the plates motion to geodesic motion can be safely neglected.

\section{The scalar field} \label{sec:ScalarField}
For the sake of simplicity we will consider a massless scalar field $\psi(x^\alpha)$ inside the cavity.
We also assume the cavity walls to be perfectly reflecting, so that the field obeys Dirichlet boundary conditions at the plates.
The generally covariant Klein-Gordon equation is \cite{birrell}
\begin{equation}\label{KG}
\frac{1}{\sqrt{-g}}\partial_\mu\big[\sqrt{-g}g^{\mu\nu}\partial_\nu\psi(x^\alpha)\big]+\kappa R(x^\beta)\psi(x^\alpha)=0,
\end{equation}
where $\kappa$ is a numerical parameter describing the coupling between the matter field and the background gravitational field and $R(x^\beta)$ is the scalar curvature.
In what follows we will suppose minimal coupling, so that $\kappa=0$.
\subsection{Tetrad form of the field equation}
The Klein-Gordon equation in the tetrad frame (\ref{tetrad}) reads \cite{karsten} \footnote{Throughout the text, a caret will mean that the corresponding quantity has to be regarded as an {\em operator}.}
\begin{equation}\label{KGtetrad}
(\Box+\hat V)\psi=0,
\end{equation}
where $\Box=\eta^{bc}\partial_b\partial_c$ is the flat d'Alembertian in the observer's Minkowski local frame and
\begin{equation}\label{V}
\hat V = \frac{1}{e}\partial_\mu (e e^\mu_{\,\, \hat a})\partial^{\hat a} = -\frac{\xi}{1-\xi\tau}\partial_\tau=b(\tau)\partial_\tau,
\end{equation}
along with
\begin{equation}
\xi=\frac{3}{2r_g},\quad\quad\quad b(\tau)=-\frac{\xi}{1-\xi\tau}.
\end{equation}
In the local frame we search for a solution obeying Dirichlet boundary conditions at the plates
\begin{equation}\label{bc}
\psi(\tau,x,\vec x_\perp)|_{x=0}=\psi(\tau,x,\vec x_\perp)|_{x=L}=0.
\end{equation}
Let us introduce, for convenience, the auxiliary field $\varphi=e^{\frac{1}{2}\int d\tau b(\tau)}\psi$, whose dynamics is the same as that of $\psi$.
Notice that $\varphi$ obeys the same boundary conditions Eq.~\eqref{bc}.
From Eq.~\eqref{KGtetrad} we get
\begin{equation}\label{eqdiff1}
\bigg[\Box +\frac{1}{4}\frac{\xi^2}{(1-\xi\tau)^2}\bigg]\varphi=0.
\end{equation}
We {\em guess} the following solution, obeying Eq.~\eqref{bc}
\begin{equation}\label{phi}
\varphi(x^a)\sim e^{i\vec k_\perp\cdot \vec x_\perp}\sin\bigg(\frac{n\pi}{L}x\bigg)\chi(\tau),\quad\quad\quad\quad n\in N
\end{equation}
where $\vec k_\perp\equiv (k_y, k_z)$, $\vec x_\perp\equiv (y, z)$, and $\chi(\tau)$ is a function of the proper (local) time, to be evaluated below.
Plugging Eq.~\eqref{phi} into Eq.~\eqref{eqdiff1} we get the following equation for $\chi(\tau)$
\begin{equation}\label{chiequation}
\bigg[\partial_\tau^2 +\omega_k^2 +\frac{1}{4}\frac{\xi^2}{(1-\xi\tau)^2}\bigg]\chi=0,
\end{equation}
where $\vec k\equiv(n\pi/L, \vec k_\perp)$ and
\begin{equation}\label{omegak}
\omega_k^2=k^2_\perp+\bigg(\frac{n\pi}{L}\bigg)^2.
\end{equation}

The dimensionless quantity $\frac{1}{1-\xi\tau}$ can be used to get an estimate of the typical rate of change of the space-time geometry surrounding the falling cavity.
If we define a time-scale
\begin{equation}
\Delta \tau=\bigg[\partial_\tau\bigg(\frac{1}{1-\xi\tau}\bigg)\bigg]^{-1},
\end{equation}
then the field modes can be considered {\em almost} stationary  by the observer if the following condition holds true
\begin{equation}\label{static}
\Delta\tau\gg\frac{1}{{\rm min}\{\omega_n\}}\simeq L.
\end{equation}
On the other hand, if $\Delta\tau\leq L$, the rate of change of the surrounding geometry is too high to assume a steady state for the field modes, and a rather different approach must be taken into account to handle a scenario in which the dynamical effects (particle creation out of the quantum vacuum) are expected to play a dominant role.

It is straightforward to check that Eq.~\eqref{static}) is satisfied in the whole time range $-\infty < \tau <0$, describing the free-fall from infinity to the black hole horizon.
Actually, in that range we have $\Delta\tau\geq\frac{1}{\xi}\simeq r_g\gg L$ by assumption.

Eq.~\eqref{chiequation} can be formally solved in terms of Bessel functions over the whole time domain $-\infty < \tau <\frac{1}{\xi}$ (from spatial infinity up to the singularity; see below).
However, beyond the horizon, the solution would become meaningless as the cavity approaches the singularity: eventually the cavity size ($L$) would become comparable with the spacetime curvature and the construction of a local frame would fail as well as the assumptions listed in Section~\ref{sec:CasimirComoving}.
To avoid such complications we will confine our analysis to the black hole exterior.

\subsection{Field mode solutions in a falling Casimir cavity}
We now need to evaluate $\chi(\tau).$ Define $\eta=1-\xi\tau$.
Then Eq.~\eqref{chiequation} becomes
\begin{equation}\label{chieq}
\frac{\partial^2\chi}{\partial\eta^2}+\bigg(\frac{\omega_k^2}{\xi^2}+\frac{1}{4\eta^2}\bigg)\chi=0,
\end{equation}
whose general solution, in terms of Bessel functions $J_0$ and $Y_0$, is
\begin{equation}
\chi_k(\eta)={\cal A}\sqrt{\eta}J_0\big(\omega_k\eta/\xi\big)+{\cal B}\sqrt{\eta}Y_0\big(\omega_k\eta/\xi\big).
\end{equation}
The choice
\begin{equation}
{\cal A}=\frac{1}{2}\sqrt{\frac{\pi}{\xi}},\quad\quad\quad {\cal B}=\frac{i}{2}\sqrt{\frac{\pi}{\xi}},
\end{equation}
yields, in terms of Hankel functions of second kind,
\begin{equation}\label{chitot}
\chi_k(\tau)=\frac{1}{2}\sqrt{\frac{\pi}{\xi}(1-\xi\tau)}H^{(1)}_0\bigg(\frac{\omega_k}{\xi}(1-\xi\tau)\bigg),
\end{equation}
which has the required minkowskian (plane wave) behaviour at $\tau\rightarrow -\infty$; when the cavity is at the spatial infinity with respect to the black hole
\begin{equation}
\chi_k(\tau)\sim \frac{1}{\sqrt{2\omega_k}}e^{-i\omega\tau},\quad\quad\quad \tau\rightarrow -\infty.
\end{equation}
The above normalized field modes Eq.~\eqref{chitot} will be used in Section~\ref{sec:subsec-BunchDavies-particle-creation} when discussing particle creation inside the cavity.

\section{Proper-time Schwinger's approach}\label{sec:SchwingerApproach}
%
%
In this section we will follow Schwinger's proper time approach \cite{schwinger1,schwinger2,cougo} in order to derive an expression of the (one-loop) effective action $W$ for the scalar field inside the Casimir cavity.
In the presence of a nonstationary gravitational background, the effective action may become {\em complex}.
In such case the real part of $W$ describes phenomena related to the vacuum polarization, as the (static) Casimir effect, and the imaginary part indicates particle production.
Actually, in the so-called  {\em in-out} formalism the imaginary part of the effective action is related to the vacuum persistence amplitude
\begin{equation}
\langle {0\, \rm out}|{0\, \rm in}\rangle=e^{i W},
\end{equation}
which in turn can be used to evaluate the number density $\langle n\rangle$ of the created field quanta.
In what follows we will evaluate both the real and the imaginary parts of the effective action.

\subsection{Computing the Effective Action}
From Eq.~\eqref{eqdiff1} the proper-time Hamiltonian $\hat H$  reads
\begin{equation}\label{H}
\hat H = \hat H_0+\hat V,
\end{equation}
where
\begin{equation}\label{H0}
\hat H_0= \partial_\tau^2-\vec\nabla^2\equiv -\hat p_0^2+\hat{\vec p^2}.
\end{equation}
As usual, we write the effective action $W$
\begin{equation}\label{Wo}
W=\lim_{\nu\rightarrow 0}W(\nu),
\end{equation}
where
\begin{equation}\label{W}
W(\nu)=-\frac{i}{2}\int_0^\infty\, ds\, s^{\nu-1}\Tr e^{-is\hat H},
\end{equation}
and the limit $\nu\rightarrow 0$ has to be taken at the end of calculations.
In Eq.~\eqref{W} the trace
\begin{equation}\label{tr}
\Tr e^{-is\hat H}=
\sumint
\,d^4x\langle x|e^{-is\hat H}|x\rangle,
\end{equation}
has to be evaluated all over the continuous as well the discrete degrees of freedom, including those of spacetime.
We write
\begin{multline}
\Tr e^{-is\hat H}=\int d^4x\,\, \sumint d\alpha\,\langle\tau,x_\perp,x|p_0,p_\perp,n\rangle\\
\times\langle p_0,p_\perp,n|e^{-is(\hat H_0+\hat V)}|p'_0,p'_\perp,n'\rangle\langle p'_0,p'_\perp,n'|\tau,x_\perp,x\rangle,
\label{tr1}
\end{multline}
where
\begin{equation}
\sumint d\alpha\,\equiv\sum_{n,n'} \int dp_0\,dp'_0\, dp_\perp dp'_\perp.
\end{equation}
Since $[\hat{\vec p},\hat V]=0$, Eq.~\eqref{tr1} can be factorized as
\begin{multline}
\Tr e^{-is\hat H}=\int d^4x\,\, \sumint d\alpha\,\langle x_\perp,x|p_\perp,n\rangle\\
\times\langle p_\perp,n|e^{is\vec\nabla^2}|p'_\perp,n'\rangle\langle p'_\perp,n'|x_\perp,x\rangle\\
\times\langle\tau | p_0\rangle\langle p_0|e^{-is\big(\partial_\tau^2+\frac{1}{4}\frac{\xi^2}{(1-\xi\tau)^2}\big)}|p'_0\rangle\langle p'_0|\tau\rangle,
\label{tr2}
\end{multline}
where
\begin{align}
    X(\vec x)& =\langle x_\perp,x| p_\perp,n\rangle\\
    T(\tau)& =\langle\tau |p_0\rangle,
\end{align}
are, respectively, the eigenfunctions of $-\vec\nabla^2$ and $\big(\partial_\tau^2+\frac{1}{4}\frac{\xi^2}{(1-\xi\tau)^2}\big)$, namely [see Eq.~\eqref{chitot}]
\begin{equation}
  \begin{split}
X(\vec x)&=\frac{1}{2\pi}\sqrt{\frac{2}{L}}e^{i\vec p_\perp\cdot\vec x_\perp}\sin\big(\frac{n\pi}{L}x\big),\\
T(\tau)&=\frac{1}{\sqrt{2\pi}}\sqrt{\frac{\pi p_0}{2\xi}(1-\xi\tau)}H^{(1)}_0\bigg(\frac{p_0}{\xi}(1-\xi\tau)\bigg).
\end{split}\label{XT}
\end{equation}
(Notice that in what follows the states $|\alpha\rangle$ are normalized according to the standard Dirac prescription: $\langle\alpha|\alpha'\rangle=\delta(\alpha,\alpha')$, where $\delta(\alpha,\alpha')$ is the Kronecker symbol $\delta_{\alpha,\alpha'}$ if  $\{|\alpha\rangle\}$ is a discrete set, and the Dirac delta function $\delta(\alpha-\alpha')$ if it is continuous).

Using Eq.~\eqref{XT} in Eq.~\eqref{tr2} and performing the $x$-integration we have
\begin{multline}
\Tr e^{-is\hat H}=\frac{1}{16\pi^2}\int d^2x_\perp d\tau\int d^2p_\perp dp_0\\
\times\sum_n e^{-is\big(p_\perp^2+(n\pi/L)^2\big)} e^{isp^2_0}\frac{p_0}{\xi}(1-\xi\tau)|H_0^{(1)}|^2,
\end{multline}
where $|H_0^{(1)}|^2=H_0^{(1)*}H_0^{(1)}$. After a $\int d^2x_\perp\, d^2p_\perp$-integration we get
\begin{multline}
\Tr e^{-is\hat H}=\frac{A}{8\pi is\xi}\int_{-\infty}^T d\tau\int_0^\infty dp_0\sum_n (1-\xi\tau) \\
\times p_0 e^{isp_0^2}\bigg|H^{(1)}_0\bigg(\frac{p_0}{\xi} (1-\xi\tau)\bigg)\bigg|^2e^{-is(n\pi/L)^2}.
\end{multline}
At any fixed $\tau$, define $q=p_0(1-\xi\tau)/\xi$.
Then
\begin{multline}
\Tr e^{-is\hat H}=\frac{A\xi}{8\pi is}\int_{-\infty}^T\frac{d\tau}{1-\xi\tau}\sum_n\int_0^\infty dq\, q \big|H^{(1)}_0(q)\big|^2\\
\times e^{-is(n\pi/L)^2} e^{is\xi^2q^2/(1-\xi\tau)^2}.
\end{multline}
Rewriting $|H^{(1)} _0(q)|^2=H_0^{(1)}(q)H_0^{(2)}(q)=J_0^2(q)+Y_0^2(q)$ and using the integral representation involving the Bessel functions $J_0$, $Y_0$ and $K_0$ \cite{glasser,grad}
\begin{multline}
J_0(a)J_0(b)+Y_0(a)Y_0(b)\\
=\frac{8}{\pi^2}\int_0^\infty\,dy\frac{\cos[(a-b)(y^2+1)^{1/2}]}{(y^2+1)^{1/2}}\, K_0[2y(ab)^{1/2}],
\label{JYK}
\end{multline}
we obtain
\begin{multline}
\Tr e^{-is\hat H}=\frac{A\xi}{\pi^3 is}\int_{-\infty}^T\frac{d\tau}{1-\xi\tau}\sum_n\int_0^\infty dq\, q \\
\times \int_0^\infty\,\frac{dy}{\sqrt{y^2+1}}K_0[2yq] e^{-is(n\pi/L)^2} e^{is\xi^2q^2/(1-\xi\tau)^2}.
\end{multline}
Performing the $q$-integration we get
\begin{multline}
\Tr e^{-is\hat H}=\frac{A\xi}{\pi^3 is}\int_{-\infty}^T\frac{d\tau}{1-\xi\tau}\sum_n  \int_0^\infty\,\frac{dy}{\sqrt{y^2+1}} \\
\times \frac{1}{4\beta} e^{iy^2/\beta}\bigg[\pi-i\Ei(-iy^2/\beta)\bigg]e^{-is(n\pi/L)^2} ,
\end{multline}
where $\Ei(z)$ is the exponential integral function and
\begin{equation}\label{beta}
\beta=\frac{s\xi^2}{(1-\xi\tau)^2}.
\end{equation}
 Performing the $y$-integration and substituting in Eq.~\eqref{W} finally yields
\begin{widetext}
\begin{equation}
W(\nu)=-\frac{iA}{32\pi^{5/2}}\int_0^\infty ds \int_{-\infty}^Td\tau\sum_n \frac{s^{\nu-3/2-1}}{\beta^{1/2}}  e^{-is(n\pi/L)^2} \bigg[\pi^{3/2}e^{-i/(2\beta)}H_0^{(1)}\big(1/(2\beta)\big)+2G_{23}^{31}\bigg(-\frac{i}{\beta}
\left |
\begin{array}{cc}
 0 &   1/2  \\
0&0\,\,\,0
\end{array}
\right)\bigg],
\label{totalW}
\end{equation}
\end{widetext}
with $G_{23}^{31}$ being a Meijer G-function.
We see that Eq.~\eqref{totalW} is made of two contributions, due to the two terms in the square brackets.
Let us consider each of them separately.

\subsection{Vacuum polarization}
The first term in Eq.~\eqref{totalW} reads
\begin{multline}
W_H(\nu)\stackrel{\rm def}{=}-\frac{iA}{32\pi^{5/2}}\int_0^\infty ds \int_{-\infty}^Td\tau\sum_n \frac{s^{\nu-3/2-1}}{\beta^{1/2}} \\
\times  e^{-is(n\pi/L)^2} \bigg[\pi^{3/2}e^{-i/(2\beta)}H_0^{(1)}\big(1/(2\beta)\big)\bigg].
\end{multline}
After expanding $H_0^{(1)}$ in powers of the dimensionless parameter $\beta$, performing some algebra, and using the Euler gamma function $\Gamma(z)=\int_0^\infty t^{z-1}e^{-t}dt$ and the Riemann zeta function $\zeta(z)=\sum_{n=1}^\infty\frac{1}{n^z}$, we obtain
\begin{multline}
W_H(\nu)=\frac{(-i)^\nu A\pi^{3/2}}{16L^3}\sum_k \xi^{2k}2^k a_k\bigg(\frac{L}{\pi}\bigg)^{2(\nu+k)}\\
\times\int_{-\infty}^T\frac{d\tau}{(1-\xi\tau)^{2k}}\Gamma(\nu-3/2+k)\zeta(2\nu-3+2k),\label{wh}
\end{multline}
where \cite{nist}
\begin{equation}
  \begin{split}
a_0&=1,\\
a_k&=\frac{1}{k!8^k}[(-1^2)(-3^2)\cdots (-(2k-1)^2)],\quad k\geq 1.
\end{split}
\end{equation}
Taking the limit $\nu\rightarrow 0$ in (\ref{wh}) we get a {\em real} quantity.

\subsection{Vacuum persistence amplitude}
Consider now the contribution to $W(\nu)$ due to the second term in the square brackets of Eq.~\eqref{totalW}.
Let us define
\begin{multline}
iW_G(\nu)\stackrel{\rm def}{=} -\frac{iA}{16\pi^{5/2}}\int_0^\infty ds \int_{-\infty}^Td\tau\sum_n\frac{s^{\nu-3/2-1}}{\beta^{1/2}} \\
\times  e^{-is(n\pi/L)^2} G_{23}^{31}\bigg(-\frac{i}{\beta}
\left |
\begin{array}{cc}
 0 &   1/2  \\
0&0\,\,\,0
\end{array}
\right).
\label{imW}
\end{multline}
Putting $\gamma=\frac{sn^2\pi^2}{\beta L^2}=\big(\frac{n\pi}{\xi L}\big)^2(1-\xi\tau)^2$, and appealing to some well-known properties of the Mejier G-functions, we rewrite $iW_G(\nu)$ as
\begin{multline}
iW_G(\nu)= -\frac{A(-i)^{\nu-3}}{16\pi^{5/2}\xi}\int_{-\infty}^T d\tau (1-\xi\tau) \sum_n\bigg(\frac{L}{n\pi}\bigg)^{2\nu-4}\\
\times G_{24}^{41}\bigg(\gamma
\left |
\begin{array}{ccc}
 \,0 &   1/2  \\
-2+\nu&0\,\,0\,\,0
\end{array}
\right)\bigg].\label{wg}
\end{multline}
Inspection of Eq.~\eqref{wg} shows that $iW_G=\lim_{\nu\rightarrow 0}iW_G(\nu)$ is an {\em imaginary} quantity.
Hence, as anticipated, we obtained a {\em complex} effective action $W$.
The real part
\begin{equation}
\realpart W=\lim_{\nu\rightarrow 0}W_H(\nu).
\end{equation}
is responsible for vacuum polarization and related phenomena, such as the static Casimir effect, as we will see in the next section.

The imaginary part reads
\begin{equation}
\imagpart W=\lim_{\nu\rightarrow 0}W_G(\nu),
\end{equation}
 implying dynamical effects, such as field quanta creation inside the cavity. We will discuss particle creation in Section~\ref{sec:DynamicalEffects}.

\section{The static Casimir Effect}\label{sec:StaticCasimir}
In this section we will discuss the {\em static} Casimir effect, deriving it from the {\em real} part of the effective action $W$.
Following Schwinger, we have from Eq.~\eqref{wh}
\begin{multline}
\langle \epsilon_\mathrm{Cas}\rangle =-\lim_{\nu\rightarrow 0}\frac{1}{AL}\frac{\partial}{\partial\tau} \realpart W(\nu) \\
=-\frac{\pi^{3/2}}{16L^4}\sum_{k=0}^\infty \frac{2^k\xi^{2k}a_k}{(1-\xi\tau)^{2k}}\bigg(\frac{L}{\pi}\bigg)^{2k}\Gamma\big(-\frac{3}{2}+k\big)\zeta(-3+2k).\label{statcas}
\end{multline}
Consider now the leading term ($k=0$) in Eq.~\eqref{statcas}.
We find
\begin{equation}
\langle \epsilon_\mathrm{Cas}\rangle^{(0)}=-\frac{\pi^{3/2}}{16L^4}a_0\Gamma(-3/2)\zeta(-3)=-\frac{\pi^2}{1440 L^4},
\end{equation}
namely the usual flat result for the Casimir energy density.
We now move to the first order correction ($k=1$) to the Casimir energy, thus obtaining
\begin{equation}
  \begin{split}
\langle \epsilon_\mathrm{Cas}\rangle^{(1)}&= -\frac{\pi^{3/2}\xi^2}{8L^2}(-1/8)\frac{1}{(1-\xi\tau)^2}\Gamma(-1/2)\zeta(-1) \\
&=\frac{\xi^2}{384L^2}\frac{1}{(1-\xi\tau)^2}.
\end{split}
\end{equation}
The Casimir energy density is then
\begin{equation}\label{mainresult}
\langle \epsilon_\mathrm{Cas}\rangle=-\frac{\pi^2}{1440L^4}+\frac{1}{384L^2}\frac{\xi^2}{(1-\xi\tau)^2}+O(\xi^4).
\end{equation}
At the horizon crossing ($\tau\rightarrow 0^-$), we have (recall that $\xi=3/(2r_g)$)
\begin{equation}
\langle \epsilon_\mathrm{Cas}\rangle_{hor}=-\frac{\pi^2}{1440L^4}\bigg[1-\frac{135}{(4\pi)^2}\bigg(\frac{L}{r_g}\bigg)^2\bigg].
\end{equation}
Eq.~(\ref{mainresult}) tells us how the corrections to the Casimir energy density change with the proper time as the cavity approaches the black hole horizon, and it holds true as long as we are in the {\em adiabatic} regime.
Namely, provided that the condition Eq.~\eqref{static} is fulfilled.

The above result shows that the comoving observer measures a small reduction in the (absolute) value of the (negative) Casimir energy near the black hole horizon. At a first glance, this may seem rather puzzling, as one would expect no change with respect to the usual flat spacetime result $\langle\epsilon_\mathrm{Cas}\rangle_{stat}=-\frac{\pi^2}{1440L^4}$ for a  {\em freely falling} Casimir cavity, due to the equivalence principle.

The resolution to this is related to other issues regarding the equivalence principle \cite{fullingjustin}.
We have implicitly assume that the cavity is prepared in the vacuum state at asymptotic infinity $\tau \rightarrow -\infty$, and we indeed see that our solution exactly recovers the flat space solution in this limit.
This state is defined on a (space-like) Cauchy surface defined by the vector field $\partial_t$ inside that cavity, but as the observer falls into the black hole, the surface with which they are observing is normal to the vector field $\partial_\tau \neq \partial_t$ (a natural consequence of the gravitational field changing with respect to proper time for the cavity while $\partial_t$ is a time-like Killing vector field).
In this sense, this \emph{pure geometric} effect we have described is an effect with memory of its (physically reasonable) initial conditions and the change of the metric along its trajectory.
This is captured by the function $T(\tau)$ in Eq.~\eqref{XT} which differs from the pure exponential usually associated with a stationary cavity.
The result is that a local measurement of $T_{\mu\nu}$ is directly related to the full unitary evolution of the initially stationary vacuum state.
The extended nature of this state allows for local observations to distinguish changing gravitational fields despite a naive application of the equivalence principle.

\section{Dynamical effects: particle creation}\label{sec:DynamicalEffects}

We have now directly alluded to the fact that we are evolving the initial vacuum state $\lvert 0\, \mathrm{in} \rangle$ with a \mbox{(proper-)time} varying gravitational field.
As such, we can explore the counterpart of the static Casimir effect: the {\em dynamical} effects induced by this time-variation (including particle creation).
The effects of looking at this in the nearly adiabatic limit will aid us by allowing us to use the same formalism as in previous sections.

\subsection{Persistence Amplitude and particle creation}
Particle creation is related to the vacuum persistence amplitude, i.e., the imaginary part of the effective action $W$.
In the in-out formalism we have
\begin{equation}
|\langle {0\, \rm out}|{0\, \rm in}\rangle|^2=e^{2i\imagpart W},
\end{equation}
where if $\lvert{0\, \rm in}\rangle$ and $\lvert{0\, \rm in}\rangle$ where unitarily related, we would have $\imagpart W = 0$, so $\imagpart W \neq 0$ indicates that the evolution of $\lvert{0\, \rm in}\rangle$ has overlap with excited states.
In fact, the (usually small) number density of created particles inside the falling cavity is
\begin{equation}
\langle n\rangle\simeq \frac{2\,\Im {\rm m}\, W}{AL}.\label{n}
\end{equation}
Consider the imaginary part Eq.~\eqref{wg} of $W$ and define $\sigma=(1-\xi\tau)^2$. We get
\begin{multline}
W_G(\nu)= \frac{A(-i)^{\nu}}{32\pi^{5/2}\xi^2}\int_{\sigma}^\infty d\sigma \sum_n\bigg(\frac{L}{n\pi}\bigg)^{2\nu-4} \\
\times G_{24}^{41}\bigg(\frac{\sigma}{\mu}
\left |
\begin{array}{ccc}
 \,0 &   1/2  \\
-2+\nu&0\,\,0\,\,0
\end{array}
\right)\bigg].\label{wg2}
\end{multline}
where $\mu=\big(\frac{\xi L}{n\pi}\big)^2$ is a small dimensionless parameter.
Upon integration we get
\begin{multline}
W_G(\nu)= \frac{A(-i)^{\nu}}{32\pi^{5/2}\xi^2} \sum_n\bigg(\frac{L}{n\pi}\bigg)^{2\nu-4}\\
\times \mu\,G_{24}^{41}\bigg(\frac{\sigma}{\mu}
\left |
\begin{array}{ccc}
 \,\,\,\,0 &\,   3/2  \\
0&1\,\,\,\,1\,\,\,\nu-1
\end{array}
\right)\bigg].\label{wgindef}
\end{multline}
The above expression is ill-defined, as the Meijer G-function Eq.~\eqref{wgindef} does not exist. However we may render it definite introducing a small quantity $\epsilon>0$, hence writing
\begin{multline}
W_G(\nu;\epsilon)= \frac{A(-i)^{\nu}}{32\pi^{5/2}\xi^2} \sum_n\bigg(\frac{L}{n\pi}\bigg)^{2\nu-4}\\
\times \mu\,G_{24}^{41}\bigg(\frac{\sigma}{\mu}
\left |
\begin{array}{ccc}
 \,\,\,\,0 &\,   3/2  \\
0+\epsilon&1\,\,\,\,1\,\,\,\nu-1
\end{array}
\right)\bigg].\label{wgepsilon}
\end{multline}
Expanding Eq.~\eqref{wgepsilon} in powers of the small parameter $\mu=\big(\frac{\xi L}{n\pi}\big)^2$ we obtain
\begin{widetext}
\begin{equation}\label{wgexp}
W_G(\nu;\epsilon)= \frac{A(-i)^{\nu}}{16\pi^{3}} \sum_n\bigg(\frac{L}{n\pi}\bigg)^{2\nu-2}\Gamma(\epsilon)
\bigg[\Gamma(\nu-1)-\frac{2\epsilon (\nu-1)\Gamma(\nu-1)}{3\sigma}\bigg(\frac{\xi L}{n\pi}\bigg)^2+\frac{8\epsilon (1+\epsilon)\nu  (\nu-1)\Gamma(\nu-1)}{15\sigma^2}\bigg(\frac{\xi L}{n\pi}\bigg)^4+\cdots\bigg]
\end{equation}
\end{widetext}
We know that in the limit $(\xi L)\rightarrow 0$ we have to recover the flat spacetime result, implying an effective action without  the imaginary part, responsible for particle creation, hence $W_G(\nu)=0$. This allows us to renormalize Eq.~\eqref{wgexp}, subtracting the divergent contribution
\begin{equation}
\lim_{(\xi L)\rightarrow 0} W_G(\nu;\epsilon)=\frac{A(-i)^{\nu}}{16\pi^{3}} \sum_n\bigg(\frac{L}{n\pi}\bigg)^{2\nu-2}\Gamma(\epsilon)
\Gamma(\nu-1).
\end{equation}
Thus, the renormalized part reads
\begin{equation}
W_G(\nu)=\lim_{\epsilon\rightarrow 0}\big[W_G(\nu;\epsilon)-\lim_{(\xi L)\rightarrow 0} W_G(\nu;\epsilon)\big].
\end{equation}
Recalling the relation $z\Gamma(z)=\Gamma(z+1)$  we have
\begin{multline}
W_G(\nu)=\frac{A(-i)^{\nu}}{24\pi^{3}} \sum_n\bigg(\frac{L}{n\pi}\bigg)^{2\nu-2}\\
\times\bigg[-\frac{\Gamma(\nu)}{\sigma}\bigg(\frac{\xi L}{n\pi}\bigg)^2+\frac{4\Gamma(\nu+1)}{5\sigma^2}\bigg(\frac{\xi L}{n\pi}\bigg)^4+\cdots\bigg].
\label{wgren}
\end{multline}
Introducing the Riemann Zeta function $\zeta(z)$ we recast Eq.~\eqref{wgren} as
\begin{multline}
W_G(\nu)= \frac{A(-i)^{\nu}}{24\pi^{2\nu+1} L^{2-2\nu}}\bigg[-\frac{\xi^2 L^2}{\pi^2\sigma}\Gamma(\nu)\zeta(2\nu) \\
+\frac{4\xi^4 L^4}{5\pi^4\sigma^2}\Gamma(\nu+1)\zeta(2\nu+2)+\cdots\bigg].
\end{multline}
Using the reflection property
\begin{equation}\label{reflection}
\Gamma\bigg(\frac{z}{2}\bigg)\zeta(z)\pi^{-z/2}=\Gamma\bigg(\frac{1-z}{2}\bigg)\zeta(1-z)\pi^{(z-1)/2},
\end{equation}
we write
\begin{multline}
W_G(\nu)= \frac{A(-i)^{\nu}}{24\pi^{7/2} L^{2-2\nu}}\bigg[-\frac{\xi^2 L^2}{\sigma}\Gamma(1/2-\nu)\zeta(1-2\nu) \\
+\frac{4\xi^4 L^4}{5\sigma^2}\Gamma(-1/2-\nu)\zeta(-1-2\nu)+\cdots\bigg].
\end{multline}
Taking the limit $\nu\rightarrow 0$ and restoring $\sigma=(1-\xi\tau)^2$ finally yields the imaginary part of the effective action $W$
\begin{equation}
\imagpart W=\frac{A}{24\pi^3 L^2}\bigg[-\frac{\xi^2 L^2}{(1-\xi\tau)^2}\zeta (1)+\frac{2\xi^4 L^4}{15(1-\xi\tau)^4}+\cdots\bigg].
\label{WGfinal}
\end{equation}
Inspection of Eq.~\eqref{WGfinal} reveals that the first term in the square brackets is still {\em divergent}.

In spite of the divergent term, when the small dimensionless quantity $\xi L= \frac{3L}{2r_g}$ is vanishing $\imagpart W\rightarrow 0$, hence implying no particle creation inside the falling cavity, as expected.
Actually, when the gravitational radius of the black hole is much greater than the plate separation, the cavity does not experience any relevant effect due to the free fall.
The persistent presence of divergences in (\ref{n}) precludes a direct evaluation of the number of created particles from the imaginary part of the effective action, unless some specific assumptions are made about the above cited infinities.
The origin of such divergence is related to the implicitly assumed {\em infinite} extension of the plates, as we will see below.
We will avoid the difficulties stemming from the appearance of infinities in the imaginary part of the effective action exploiting the relationship between the Schwinger theory and the in-out formalism, based upon the Bogolubov  approach.

\subsection{Bunch-Davies vacuum and particle creation}\label{sec:subsec-BunchDavies-particle-creation}
Recall the field modes Eq.~\eqref{chitot} we found in Section~\ref{sec:ScalarField}
\begin{equation}\label{chitot2}
\chi_k(\eta)=\frac{1}{2}\sqrt{\frac{\pi}{\xi}\eta}H^{(1)}_0\bigg(\frac{\omega_k}{\xi}\eta\bigg),\quad\quad\quad \eta=1-\xi\tau,
\end{equation}
which have the required minkowskian (plane wave) behaviour at $\eta\rightarrow \infty$ (i.e. $\tau\rightarrow -\infty$) when the cavity is at spatial infinity with respect to the black hole.
The above modes satisfy the Bunch-Davies vacuum requirements, namely
\begin{equation}
\left.
\begin{array}{ll}
\chi_k(\eta)\rightarrow \frac{1}{\sqrt{2\omega_k}}e^{i\frac{\omega_k}{\xi}\eta}\sim \frac{1}{\sqrt{2\omega_k}}e^{-i\omega_k\tau}\nonumber\\
\frac{\dot\chi_k(\eta)}{\chi_k(\eta)}\rightarrow i\frac{\omega_k}{\xi}
\end{array}
\right\} \quad\quad \eta\rightarrow \infty.
\end{equation}
Also,  we see that as far as
\begin{equation}\label{etacondition}
\eta\gg\frac{\xi}{2\omega_k},
\end{equation}
Eq.~\eqref{chieq} reduces to
\begin{equation}\label{chieq2}
\frac{\partial^2\chi}{\partial\eta^2}+\bigg(\frac{\omega_k^2}{\xi^2}\bigg)\chi=0,
\end{equation}
so, in the far past Eq.~\eqref{chieq2} admits a plane wave solution
\begin{equation}\label{chiplane}
\chi_k(\tau)=\frac{\alpha}{\sqrt{2\omega_k}}e^{-i\omega_k\tau}+\frac{\beta}{\sqrt{2\omega_k}}e^{i\omega_k\tau}.
\end{equation}
From Eq.~\eqref{etacondition} we get
\begin{equation}
\eta\gg\frac{\xi}{2\omega_k}=\frac{3}{4\omega_k r_g}.
\end{equation}
Since $\omega^2_k=k^2_\perp + (n\pi/L)^2$,  we have $\min(\omega_k)=\pi/L$.
So, if $\eta\gg\frac{\xi L}{2\pi}$, then Eq.~\eqref{etacondition} is undoubtedly fulfilled.
Obviously, at the horizon crossing  $\eta =1\gg\frac{\xi L}{2\pi}$, so any point near the horizon, characterized by $\eta\geq 1$, can be used to match the solutions Eq.~\eqref{chitot} and Eq.~\eqref{chiplane} by demanding that both $\chi_k$ and $\partial\chi_k/\partial\tau$ are continuous at the chosen boundary $\eta\geq 1$, namely
\begin{widetext}
\begin{equation}\begin{split}
\frac{\alpha}{\sqrt{2\omega_k}}e^{-i\omega_k\tau}+\frac{\beta}{\sqrt{2\omega_k}}e^{i\omega_k\tau} &= \frac{1}{2}\sqrt{\frac{\pi}{\xi}(1-\xi\tau)}H^{(1)}_0\bigg(\frac{\omega_k}{\xi}(1-\xi\tau)\bigg),\\
\frac{-i\omega_k\alpha}{\sqrt{2\omega_k}}e^{-i\omega_k\tau}+\frac{i\omega_k\beta}{\sqrt{2\omega_k}}e^{i\omega_k\tau} &= \frac{1}{2}\sqrt{\frac{\pi}{\xi}}\bigg[\frac{-\xi}{2\sqrt{(1-\xi\tau)}}H^{(1)}_0\bigg(\frac{\omega_k}{\xi}(1-\xi\tau)\bigg)+\sqrt{1-\xi\tau}\omega_k H^{(1)}_1\bigg(\frac{\omega_k}{\xi}(1-\xi\tau)\bigg)\bigg].
\label{matching}
\end{split}\end{equation}
\end{widetext}
After some algebra, we get the Bogolubov coefficients \footnote{(An interesting approach, requiring no detailed knowledge of state normalization, based upon the paper by Hamilton et al. (A. Hamilton, D. Kabat and M. Parikh, {\it JHEP} {\bf 0407}, 024 (2004)), may be used as well to obtain the same result).}
\begin{align}\label{bog}
|\alpha_k|^2& =1+\frac{\xi^2}{16\omega_k^2 (1-\xi\tau)^2},\\
|\beta_k|^2& =\frac{\xi^2}{16\omega_k^2 (1-\xi\tau)^2},
\end{align}
satisfying $|\alpha_k|^2-|\beta_k|^2=1$.
The $\beta$ coefficient is related to particle creation.
Note that, as $\tau\rightarrow -\infty$,  $|\alpha_k|^2\sim 1$ and $|\beta_k|^2\sim 0$, i.e., we have no particle creation in the far past, as expected, meanwhile at the horizon crossing ($\tau =0$)  we have $|\beta_k|^2=\frac{\xi^2}{16\omega_k^2}$.
We evaluate the density of created quanta as
\begin{equation}
  \begin{split}
\langle n\rangle=\frac{1}{AL}\bigg[\frac{A}{(2\pi)^2}\sum_n\int d^2 k_\perp\frac{\xi^2}{16\omega_k^2\eta^2}\bigg]\\
=\frac{\xi^2}{64\pi^2 L\eta^2}\sum_n \int \frac{d^2 k_\perp}{k^2_\perp +(n\pi/L)^2}.
\label{particledensity}
\end{split}
\end{equation}
 Using
\begin{equation}\label{ktool}
\int \frac{d^2 k_\perp}{(k^2_\perp +\sigma)^\alpha}=\pi\frac{\Gamma(\alpha-1)}{\Gamma(\alpha)}\frac{1}{\sigma^{\alpha -1}},
\end{equation}
we obtain
\begin{equation}
\langle n\rangle=\frac{\xi^2 \gamma(3/2-\alpha)\zeta(3-2\alpha)}{64 \pi^{3/2}L^{3-2\alpha}}\stackrel{\alpha=1}{\longrightarrow}\frac{\xi^2}{64 \pi L}\zeta(1),
\end{equation}
namely a {\em divergent} result.
This basically agrees with the divergent quantity we found in the imaginary part of the effective action.
Now we see that the divergence appears as a consequence of the $k_\perp$-integration over the transverse modes of the quantum field.
This implicitly involves an infinite transverse (hence unphysical) extension of the cavity.

In spite of the above divergent result, we can get a {\em finite} result for the energy density $\langle \epsilon_{\rm dyn}\rangle$  of the created quanta, writing
\begin{equation}
  \begin{split}
\langle \epsilon_{\rm dyn}\rangle=&\frac{1}{AL}\bigg[\frac{A}{(2\pi)^2}\sum_n\int d^2 k_\perp\frac{\xi^2}{16\omega_k^2\eta^2}\omega_k\bigg] \\
&=\frac{\xi^2}{64\pi^2 L\eta^2}\sum_n \int \frac{d^2 k_\perp}{\big(k^2_\perp +(n\pi/L)^2\big)^{1/2}}.
\label{particleenergy}
\end{split}
\end{equation}
Using again Eq.~\eqref{ktool} we obtain
\begin{equation}
\langle \epsilon_{\rm dyn}\rangle=-\frac{\xi^2}{32L^2\eta^2}\zeta(-1)=\frac{\xi^2}{384L^2(1-\xi\tau)^2}.
\label{DCE}
\end{equation}
If we compare the above result with Eq.~\eqref{mainresult}, describing the vacuum energy density pertaining to the Casimir effect
\begin{equation}\label{mainresult2}
\langle \epsilon_\mathrm{Cas}\rangle=-\frac{\pi^2}{1440L^4}+\frac{1}{384L^2}\frac{\xi^2}{(1-\xi\tau)^2}.
\end{equation}
we see, quite interestingly, that the small reduction observed in the static Casimir energy value {\em exactly} corresponds to the amount of energy of created field particles.
This could suggest a relationship between the two considered effects.
Nevertheless, some care is required when speculating about such coincidence, as both results have been obtained as first-order approximations.

\section{Discussion}\label{sec:Discussion}
We are now in a position to draw some conclusions about Casimir effect inside a small cavity, freely falling into a Schwarzschild black hole, with particular concern in the late stages of the fall.

Comparison of Eq.~\eqref{DCE} and Eq.~\eqref{mainresult2} shows that the overall energy density (as measured by the comoving observer) can be considered as made of two contributions
\begin{itemize}
\item{The first one, related to the vacuum polarization, is the {\em static} Casimir effect contribution, $\langle\epsilon_\mathrm{Cas}\rangle$, whose minkowskian value $\langle\epsilon_\mathrm{Cas}\rangle_0=-\frac{\pi^2}{1440L^4}$ has been modified by a small (positive) term, due to the cavity fall.}

\item{The second one, $\langle\epsilon_{\rm dyn}\rangle=\frac{1}{384L^2}\frac{\xi^2}{(1-\xi\tau)^2}$, is related to the vacuum persistence. It represents a {\em dynamical}  contribution, due to the time dependent background experienced by the quantum field, leading to particle creation inside the Casimir cavity.}
\end{itemize}

At a first glance, one could wonder whether corrections to the static Casimir effect as well as particle creation are detected by an observer in a freely falling {\em inertial} frame.
However, as anticipated at the end of Section~\ref{sec:StaticCasimir}, this is not so surprising.
The equivalence principle (EP), deeply rooted in the theory of General Relativity (GR), applies well in the context of a {\em local} theory, just as GR is.
On the other hand, when quantum fields are taken into account, the state requires definition on an entire spacelike Cauchy surface which can lead to effects seemingly in conflict with the EP, causing the latter to be not straightforwardly applicable.

In the present scenario, the quantum field stress-energy tensor $T_{\mu\nu}$ probes the history and extent of the full quantum field, thus probing the full spacetime structure between the plates as it evolves, through the long wavelength field modes (to be clear, causality is never violated by measuring this object).
The adopted renormalization procedure (whatever it may be) helps to establish the full quantum evolution and as such transfers the spacetime details into the renormalized $T_{\mu\nu}^{\rm ren}$, which is the locally {\em measured} object in this work.
In such a way, information contained in the changing spacetime geometry surrounding the cavity can be probed despite EP (in a manner similar to how classical fluids can seemingly violate EP by the observation of tidal forces). These corrects appear both in the form of a small correction to the expected static Casimir energy and a tiny flux of created field quanta.

\section{Concluding Remarks}\label{sec:Conclusions}

In this paper we have considered the Casimir energy density corrections in a small cavity freely falling from the spatial infinity into a Schwarzschild black hole. The main results of the present work are Eqs.~\eqref{mainresult2} and \eqref{DCE} representing, respectively, the (static) Casimir energy density and the energy density due to creation of field quanta inside the cavity.

As discussed above, particle creation in an inertial (freely falling) physical system could be justified recalling that the geometry seen by the cavity changes dynamically throughout its fall, and this is imprinted on $T_{\mu\nu}$ whose renormalized part is ultimately the object which is measured by the comoving observer.

In deriving the above results several assumptions have been made (see Section~\ref{sec:CasimirComoving}).
In particular, we have neglected other possible contributions deriving from the cavity extension. Tidal effects, for example, are expected to give rise to anisotropies in the energy density distribution inside the cavity; such aspect has been considered in detail in paper \cite{justin}, working out a 1+1D model.

Also, the finiteness of the Casimir plates has not been taken into account, assuming $L\ll\sqrt{A}\ll r_g$.
Such assumption is obviously fulfilled in any realistic scenario where the gravitational radius of a black hole is undoubtedly many order of magnitude greater than the cavity size.
One could  think as well of a micro-black hole, having a gravitational radius $r_g\sim L$.
But, in such a case the above equations would become meaningless (the condition $L/r_g\ll 1$ is violated), as in that limit  the local frame couldn't be considered {\em almost} minkowskian (the tidal effects will dominate).

Another drawback stemming from the (in)finiteness of the Casimir plates  appears in the divergences we met in evaluating the number density of the created quanta (both working with the effective action and the in-out formalism).
We have seen that the appearance of this divergence is related to the extensive number of modes in the transverse direction, and as such any truly finite size $A$ of the plates would cause this number to be finite (but scale with $A$).

We wish to point out that, exploiting the Lema\^itre coordinates (well-behaved up to the singularity), it could be interesting  (although not so straightforward), to explore the dynamics of the Casimir energy (according to the comoving observer) in the region $0<\tau<\frac{2r_g}{3}$, corresponding to the proper time lapse required to reach the central singularity.

In that respect, the adopted Schwinger approach seems of particular interest, as it might represent a starting point for a deeper non-perturbative analysis.

An obvious improvement of the present research would be to extend the analysis of \cite{justin} to the 3+1D case,  also including both {\em tidal} and {\em 3D-finite-size} effects in evaluating the corrections to the Casimir effect. We leave these extensions to future work.

\begin{acknowledgments}
We thank Steve Fulling for collaborations on related work and helpful discussions. We also thank Yoni Bentov for helpful discussions.
This work was supported by the Caltech Institute for Quantum Information and Matter, an NSF Physics Frontiers Center with support of the Gordon and Betty Moore Foundation and the Air Force Office for Scientific Research (J.H.W.).
\end{acknowledgments}

\end{document}